\documentclass[11pt,notitlepage,twoside,a4paper]{article}

\usepackage{hyperref}
\usepackage{listings}

\newcommand{\ignore}[1]{}

\begin{document}

\author{Tommaso Urli \thanks{\emph{Scheduling and Time-Tabling Group}, DIEGM - University of Udine, Via delle Scienze 206, 33100 -- Udine (UD), Italy}\\\texttt{\href{mailto:tommaso.urli@uniud.it}{tommaso.urli@uniud.it}}}
\title{\textsc{Technical Report}\\json2run: a tool for experiment design \&  analysis}

\maketitle

\tableofcontents

\section{Introduction}

\textbf{json2run} is a tool to automate the running, storage and
analysis of experiments. It has been created in the first place to study
different algorithms or different sets of values for algorithm
parameters, but it is a general tool and can be used wherever it fits.
The main advantage of \textbf{json2run} (over a home-brewed experiment
suite) is that it allows to describe a set of experiments concisely as a
\href{http://www.json.org}{JSON}-formatted parameter tree, such as the
following (note the presence of parameter definitions as well as logical
operators to combine them)

\begin{scriptsize}
\begin{lstlisting}
{
    "type": "and",
    "descendants": [
        {
            "type": "discrete",
            "name": "a",
            "values": [ "foo", "bar", "baz" ]
        },
        {
            "type": "or",
            "descendants": [
                {
                    "type": "discrete",
                    "name": "b1",
                    "values": { "min": 0.0, "max": 1.0, "step": 0.25 }
                },
                {
                    "type": "discrete",
                    "name": "b2",
                    "values": { "min": 2.0, "max": 10.0, "step": 2.0 }
                }
            ]
        }
    ]
}
\end{lstlisting}
\end{scriptsize}

An experiment file such as the one above describes the parameters that
must be generated and passed over to the executable. We'll call a set of
generated parameters a \emph{configuration} or \emph{parameter
configuration}. Once the experiments have been described,
\textbf{json2run} can parse the file and perform various operations with
it, such as

\begin{itemize}
\item
  printing the generated configurations as command line options,
\item
  running a batch of experiments based on the generated configurations,
\item
  store the results of the experiments in a database,
\item
  running a parameter race (see \cite{Birattari2010}) to find out the configurations (or
  configuration) that optimize a function of quality,
\item
  retrieving the results of the experiments from the database.
\end{itemize}

In the first case, the outcome of our above example would be something
like this (\textbf{json2run} comes in form of a command line tool called
\texttt{j2r}):

\begin{lstlisting}
$ j2r -i experiments.json
--a foo --b1 0.0
--a foo --b1 0.25
...
--a baz --b2 8.0
--a baz --b2 10.0
\end{lstlisting}

\textbf{json2run} supports a number of different types of nodes in the
JSON tree, including: nodes for generating parameters from a discrete
set of variables, nodes for generating parameters from the content of a
directory or a file, nodes for sampling values from an interval, and so
on. If something cannot be expressed with simple parameter generators, a
number of \textbf{post-processors} allow you to mix, merge and discard
the generated parameters in extremely flexible ways. Most
post-processors were created because something couldn't be expressed
with simple logical operators, but \textbf{json2run} slowly converged to
something complete now.

The experiments results are stored on a (MongoDB) database, in order to
be accessed later for analysis. The choice of MongoDB comes from the
necessity of comparing algorithms which can have different (and a
different number of) parameters, and a using a tabular storage (as in
most relational database) would make queries much more difficult.

Finally, \textbf{json2run} comes with a very general but handy R script,
which allows to gather data from the database, and do whatever kind of
statistical analysis over it.

\section{Installation}

Being packaged as a python module, the installation of \textbf{json2run}
should be quite straightforward, just ensure that the \texttt{bson}
python module is not installed on your system (if this is the case, run
\texttt{sudo pip uninstall bson}) since \texttt{pymongo} comes with its
own \texttt{bson.*} classes and conflicts may occur. Then, clone the
(Mercurial) repository and run the python installer

\begin{small}
\begin{lstlisting}
hg clone https://tunnuz@bitbucket.org/tunnuz/json2run
cd json2run
sudo python setup.py install
\end{lstlisting}
\end{small}

\noindent
or, if you plan to update \textbf{json2run} often (i.e.~if you plan to
customize it, or update it from the repository), run

\begin{lstlisting}
sudo python setup.py develop
\end{lstlisting}

\noindent
this will allow you to update to the latest version by just running

\begin{lstlisting}
hg pull -u
\end{lstlisting}

\noindent
in the root directory.

\section{Usage}

Since \textbf{json2run} is designed to be very flexible (its only
requirements being that you expose all the parameters of your executable
and that you have access to a MongoDB instance), this also means that it
comes with a lot of options. We will go through them in the following
sections, but if you just need a quick reference type

\begin{lstlisting}
$ j2r --help
\end{lstlisting}

Before running anything, however, you will need to know how to write an
experiment file.

\subsection{Designing experiment files}

As previously mentioned, \textbf{json2run} expects a description of the
experiments in JSON format. JSON (which, by the way, stands for
JavaScript Object Notation) is a concise and human-readable language for
describing structured data. It is also the very language MongoDB uses to
store its data and to make queries, which makes for a natural
integration.

\subsubsection{Basic JSON syntax}

The basic components of a JSON documents are arrays, objects and
scalars. You can use array and object to group more arrays and objects,
or scalars.

\paragraph{Scalars}
\noindent
JSON has a number of native types for scalars:

\begin{itemize}
\item
  numbers,
\item
  strings, and
\item
  booleans.
\end{itemize}
Numbers can be integers or floats, and scientific notation is also
supported.

\paragraph{Arrays and objects}
Arrays are lists of scalars separated by commas, e.g.:

\begin{lstlisting} 
[1, 2, 3, "foo", 3.14, true]
\end{lstlisting}

\noindent
Objects can be seen as named arrays (similar to C++'s map, or
dictionaries if you're into Python):

\begin{lstlisting} 
{  
    "name": "John",
    "surname": "Boags",
    "profession": "beer maker",
    "age": 150
}
\end{lstlisting}

\noindent
Note that arrays and objects use a different kind of parenthesis,
\texttt{\{\}} vs \texttt{{[}{]}}.

\paragraph{Comments}

JSON doesn't support comments, and most of the time you won't need them
(the JSON contents should be sufficiently explanatory), but if you
really need annotations, you can add fake entries to your objects, that
won't be parsed by \textbf{json2run} and will serve as comments, such as
\texttt{comment} in the following example:

\begin{small}
\begin{lstlisting} 
{ 
    "type": "discrete", 
    "name": "initial_temperature", 
    "values": { "min": 10, "max": 30, "step": 10 }
    "comment": "The initial temperature for SA."
}
\end{lstlisting}
\end{small}

\noindent
\paragraph{Specific syntax}
In particular, \textbf{json2run} assumes that the experiment file is a
representation of a tree in which each node is a JSON object with at
least a \texttt{type} field describing its type, e.g.:

\begin{lstlisting} 
{
    "type": "<node_type>",
    ... 
}
\end{lstlisting}

\noindent
We'll see the supported node types and their additional fields in the
following sections.

\subsubsection{Combinations and alternatives (inner nodes)}

Typically, an algorithm accepts multiple parameters, and we want to be
able to compare alternative combinations of these parameters.
\texttt{and} and \texttt{or} nodes are the way to accomplish this in
\textbf{json2run} and we call them \emph{inner} nodes. Each inner node
has a list of descendants and when they are activated, they either
combine them together (in case of an \texttt{and}), or pick between them
(in case of an \texttt{or}).

\paragraph{\texttt{and} nodes}

For instance, suppose that we have a Simulated Annealing \cite{Kirkpatrick1983} solver that accepts an initial temperature and a
cooling schedule as command line parameters, and we would like to try
all the possible combinations. In \textbf{json2run} this is expressed
using an \texttt{and} node that combines the two parameters (ignore for
now the syntax to describe discrete parameters, we'll come to that
later).

\begin{footnotesize}
\begin{lstlisting}
{
    "type": "and",
    "descendants": [
        { 
            "type": "discrete", 
            "name": "initial_temperature", 
            "values": { "min": 10, "max": 30, "step": 10 }
        },
        {
            "type": "discrete",
            "name": "cooling_schedule",
            "values": [ 0.999, 0.99, 0.9 ]
        }
    ]
}

$ j2r -i experiments.json
--initial_temperature 10.0 --cooling_schedule 0.999
--initial_temperature 10.0 --cooling_schedule 0.99
--initial_temperature 10.0 --cooling_schedule 0.9
--initial_temperature 20.0 --cooling_schedule 0.999
--initial_temperature 20.0 --cooling_schedule 0.99
--initial_temperature 20.0 --cooling_schedule 0.9
--initial_temperature 30.0 --cooling_schedule 0.999
--initial_temperature 30.0 --cooling_schedule 0.99
--initial_temperature 30.0 --cooling_schedule 0.9
\end{lstlisting}
\end{footnotesize}

\paragraph{\texttt{or} nodes}

Sometimes you have algorithms that accept different parameters and,
possibly, a different number of them. Suppose your solver can operate
either using Simulated Annealing and Tabu Search. You can easily encode
this in an experiment file by using an \texttt{or} node.

\begin{scriptsize}
\begin{lstlisting}
{
    "type": "or",
    "descendants": [
        {
            "type": "and",
            "descendants": [
                { 
                    "type": "discrete", 
                    "name": "algorithm", 
                    "values": [ "sa" ]
                },
                { 
                    "type": "discrete", 
                    "name": "initial_temperature", 
                    "values": { "min": 10, "max": 20, "step": 10 }
                },
                {
                    "type": "discrete",
                    "name": "cooling_schedule",
                    "values": [ 0.999, 0.99 ]
                }
            ]
        },
        {
            "type": "and",
            "descendants": [
                { 
                    "type": "discrete", 
                    "name": "algorithm", 
                    "values": [ "ts" ]
                },
                { 
                    "type": "discrete", 
                    "name": "tabu_list_length", 
                    "values": [ 10, 15, 20 ]
                }
            ]
        }
    ]
}
\end{lstlisting}

\begin{lstlisting}
$ j2r -i experiments.json
--algorithm sa --initial_temperature 10.0 --cooling_schedule 0.999
--algorithm sa --initial_temperature 10.0 --cooling_schedule 0.99
--algorithm sa --initial_temperature 20.0 --cooling_schedule 0.999
--algorithm sa --initial_temperature 20.0 --cooling_schedule 0.99
--algorithm ts --tabu_list_length 10
--algorithm ts --tabu_list_length 15
--algorithm ts --tabu_list_length 20
\end{lstlisting}
\end{scriptsize}

Note that no Tabu Search parameters appear in the Simulated Annealing
configurations, and vice versa. By combining several \texttt{and} and
\texttt{or} node, it is possible to express quite complex experiment
designs.

\subsubsection{Leaf nodes}

Leaf nodes are responsible for creating values for single parameters.
They all come with a \texttt{type}, a \texttt{name} for the parameter,
and an array (or object) of \texttt{values} describing the possible
values that the parameter can take, e.g.:

\begin{lstlisting}
{
    "type": "<leaf_type>",
    "name": "<parameter_name>",
    "values": <values_definition>
}
\end{lstlisting}

\paragraph{\texttt{discrete} nodes}

Discrete nodes are the simplest kind of leaf nodes. They come with two
value definition styles: an \emph{explicit} one, where parameters are
listed explicitly, e.g.:

\begin{lstlisting}
{
    "type": "discrete",
    "name": "num_of_reheats",
    "values": [ 1, 2 ]
}
\end{lstlisting}

\noindent
and an \emph{implicit} one, which allows to define discrete numeric
values from a \texttt{min}, a \texttt{max} and a \texttt{step}, e.g.:

\begin{small}
\begin{lstlisting}
{
    "type": "discrete",
    "name": "start_temperature",
    "values": { "min": 0.0, "max": 10.0, "step": 0.034 }
}
\end{lstlisting}
\end{small}

During the processing of the tree, implicit value definitions are
transformed into explicit ones, and treated as such. A \texttt{discrete}
node generates all the possible parameter values in order.

\paragraph{\texttt{continuous} nodes}

Continuous nodes allow to define parameter values that are generated by
sampling continuous parameter spaces. These spaces are defined in terms
of a \texttt{min} and a \texttt{max}, e.g.:

\begin{lstlisting}
{
    "type": "continuous",
    "name": "start_temperature",
    "values": { "min": 0.0, "max": 10.0 }
}
\end{lstlisting}

However, continuous nodes can't generate parameter values by themselves.
Instead, they need to be processed later on by a \emph{post-processor}
attached to a node upper in the tree hierarchy. This might seem over
complicated, but there's a use case behind it.

In particular suppose that you want to study the interaction of two
parameters on the performance of an algorithm. To do a proper sampling,
the generated parameters must be picked from a 2-dimensional space, in a
way that is as uniform a possible. One way to do this, would be to
generate an \texttt{and} node containing several \texttt{discrete} nodes
with different ranges. There are two problems with this approach (which
is called \emph{full-factorial}):

\begin{enumerate}
\item
  the generated points are very regular, while one usually want to
  sample the parameter space randomly (but uniformly),
\item
  the parameter combinations are the carthesian product of the values
  generated for each single parameter, which makes it difficult to
  control how much configurations are generated.
\end{enumerate}

While this can still be done with \textbf{json2run} (and indeed is often
what one wants), we would like to treat the parameter space generated by
the interaction of the two parameters as a \emph{single} space, and
sample 2-dimensional points uniformly inside it. \textbf{json2run} comes
with a post-processor which is able to generate the Hammersley point set
in a k-dimensional space. We will see in the section about
post-processors how to attach one to a node, but for the moment just
accept that \texttt{continuous} leaf nodes are treated in this special
way.

\paragraph{\texttt{file} and \texttt{directory} nodes}

File (or directory) nodes are essentially \texttt{discrete} nodes, whose
values are nor defined explicitly nor implicitly, but instead are
generated from the content of a file (or a directory). The typical use
for this kind of nodes is to generate experiments that run on a set of
instances specified in a file e.g.:

\begin{lstlisting}
{
    "type": "file",
    "name": "instance",
    "path": "selected_instances.txt",
    "match": ".*"
}
\end{lstlisting}

\noindent
where the \texttt{path} field specifies the location of the file to be
used as input, and the \texttt{match} field restrict the generated
parameters to the lines of the file matching a given regular
expression\footnote{\textbf{Note} regular expressions are in Python
  format, but strings must be escaped, e.g.~if you want to look for the
  .txt pattern, the string must be specified as
  ``.*\textbackslash{}\textbackslash{}.txt''}. This is useful when you
want to restrict to certain instances, but most frequently will just be
``.*'' (catch-all). As for directory nodes, they follow a similar
semantic, the difference being that the generated values is the list of
the content of the directory, filtered by the regular expression in the
\texttt{match} field.

\begin{lstlisting}
{
    "type":"directory",
    "path": "../instances/comp",
    "name": "instance",
    "match": ".*\\.ectt"
}
\end{lstlisting}

\paragraph{\texttt{flag} nodes}

Flag nodes have a single parameter (the \texttt{name} of the generated
flag) and generate value-less parameters. E.g.:

\begin{lstlisting}
{
    "type": "flag",
    "name": "verbose"
} 
\end{lstlisting}

\noindent
will just add the \texttt{-{}-verbose} flag on the generated command
lines.

\subsubsection{Post-processors}

Post-processors are tools to generate more complex combinations of
parameters. They can be attached to \textbf{any} inner node by adding
them to a field called \texttt{postprocessors} along with the
descendants, e.g.:

\begin{lstlisting}
{
    "type": "and",
    "descendants": [
        ...  
    ],
    "postprocessors": [
        ...
    ]
}
\end{lstlisting}

Post-processors have a \texttt{type} field and a number of other fields
dependent on the specific post-processor type. They all operate in the
following way:

\begin{enumerate}
\item
  They take the list of parameters generated in the subtree (note that
  each execution of a subtree gives birth to a different parameter
  configuration) they are attached to,
\item
  they process the list \textbf{as a whole} (e.g.~replacing parameters,
  modifying values, removing parameters, and so forth), and finally
\item
  they return the new list, which replaces the old one.
\end{enumerate}

In many cases they just apply the same function over all the elements of
the list, but they might be designed to do more complex things or to
just update certain kind of parameters. For instance, the
\texttt{hammersley} post-processor only apply to \texttt{continuous}
parameters (but possibly more than one of them at a time).

\textbf{Note} order matters! Post-processors are applied in the order in
which they are defined in the \texttt{postprocessors} list.

\paragraph{\texttt{expression} processors}

Expression post processors are by far the most flexible ones. They allow
to define a new parameter (either \texttt{discrete} or
\texttt{continuous}, but \textbf{not} a flag) by evaluating a python
expression and using the result as the value of the parameter. The
processor is defined by a \texttt{match} regular expression, which
captures the operands needed by the expression, and either

\begin{enumerate}
\item
  an \texttt{expression} which will be evaluated to yield the value of
  the generated \texttt{discrete} parameter, or
\item
  two expressions (\texttt{min} and \texttt{max}) that will yield the
  values for the minimum and maximum of the generated
  \texttt{continuous} parameter.
\end{enumerate}

The type of the parameter is inferred by the presence of the
\texttt{expression} field, while its name is defined by the
\texttt{result} field. An example of the two syntaxes is the following:

\begin{lstlisting}
{
    "type": "expression",
    "match": "<operand_1>|<operand_2>|..."
    "result": "<parameter_name>",
    "expression": "<expression>"
}

{
    "type": "expression",
    "match": "<operand_1>|<operand_2>|...",
    "result": "<parameter_name>",
    "min": "<min_expression>",
    "max": "<max_expression>"
}
\end{lstlisting}

\paragraph{Expression syntax} Any valid python expression that has a return value can be used as
\texttt{expression}, \texttt{min} or \texttt{max}. To access the values
of the captured operands it is sufficient to postfix their name with
\texttt{.value}.

As for the available operations, functions from python's \texttt{math}
and \texttt{json} modules are automatically imported. For instance, to
generate a new parameter \emph{p3} which is the power of two existing
ones, \emph{p1} and \emph{p2}, we'll write:

\begin{lstlisting}
{
    "type": "expression",
    "match": "p1|p2",
    "result": "p3",
    "expression": "pow(p1.value, p2.value)"
}
\end{lstlisting}

\newpage
\noindent
While to generate a parameter which takes values in \emph{{[}0.1*p1,
5*p2{]}}, we'll do:

\begin{lstlisting}
{
    "type": "expression",
    "match": "p1|p2",
    "result": "p3",
    "min": "0.1*p1.value",
    "max": "5*p2.value"
}
\end{lstlisting}

\paragraph{\texttt{ignore} processors}

Ignore post processors can be used to remove specific parameters from
the list of generated ones. Typically, the are used to discard operands
of an \texttt{expression} post-processor after they have been used.
Following the previous example, we might not be interested in \emph{p1}
and \emph{p2} at all, so:

\begin{lstlisting}
{
    "type": "ignore",
    "match": "p1|p2"
}
\end{lstlisting}

Remember that post-processors are applied in order, thus (in this case)
the \texttt{ignore} must be defined after the \texttt{expression}.

\paragraph{\texttt{sorting} processors}

Sorting allows to define an ordering for a subset of parameters. These
parameters will be put (if they exist) at the beginning of the generated
list of parameters, and the others will follow. The syntax of the
post-processor is the following:

\begin{lstlisting}
{
    "type": "sorting",
    "order": [ "<param_1>", "<param_2>", ... ]
}
\end{lstlisting}

\noindent
Where \texttt{order} is an array of ordered parameter names.

\paragraph{\texttt{hammersley} processors}

The Hammersley post-processor generates the scaled k-dimensional
Hammersley point set from a set of k \texttt{continuous} parameters, and
it's the preferential (also, the only) way to sample continuous
parameter spaces. The syntax is the following:

\begin{lstlisting}
{
    "type": "hammersley",
    "points": <n>
}
\end{lstlisting}

So, assuming that your experiments file produces \emph{k} continuous
parameters, the \texttt{hammersley} post-processor generates a
k-dimensional \emph{cube} delimited by the \texttt{min} and \texttt{max}
fields of your \texttt{continuous} parameters, and will generate
\texttt{n} samples inside this cube, to use as parameter values.

\paragraph{The Hammersley point set} This choice of the Hammersley point set has been driven by two
properties of this sequence that make it favourable to parameter tuning.
First, points from the Hammersley set exhibit \emph{low discrepancy},
i.e.~they are well distributed across the parameter space despite being
random-like. Second, the sequence is \emph{scalable} both with respect
to the number of points (\texttt{n}) to be sampled and to the number of
dimensions (\texttt{k}) of the sampling space.

So, whenever you want to explore parameter spaces, use
\texttt{continuous} parameters and the \texttt{hammersley}
post-processor.

\paragraph{\texttt{rounding} processors}

When sampling continuous parameters or using \texttt{expression}
post-processors, the resulting values can end up being floats with many
decimal digits. While this in general is not an issue, often this much
precision is unneeded, and it's just more convenient to operate with
less precise floats. The \texttt{rounding} post-processor allows to
round down a parameter's values to a specific number of decimal digits.
The syntax is the following:

\begin{lstlisting}
{
    "type": "rounding",
    "match": "<regex>"
    "decimal_digits": <n>
}
\end{lstlisting}

\noindent
Where \texttt{n} is the number of decimal digits we want to retain
(\textbf{note} the numbers are rounded, not truncated, to \texttt{n}
digits after the floating point), and \texttt{match} is a regular
expression describing the parameters we want to round down.

\paragraph{Compact syntax} Rounding post-processors also support a compact syntax, to group
roundings in a single post-processor. The syntax is the following:

\begin{lstlisting}
{
    "type": "round",
    "round": [
        "<regex_1>": <n_1>,
        "<regex_2>": <n_2>,
        ...
    ]
}
\end{lstlisting}

\noindent
Where \texttt{regex\_k} are regular expressions describing one (or more)
parameter, and \texttt{n\_k} are the corresponding decimal digits we
want to retain.

\paragraph{Forcing precision} Both syntaxes support an optional field called
\texttt{force\_precision}, which can be either \texttt{true} or
\texttt{false}, that forces the resulting value to have the specified
number of decimal digits (regardless of any rounding to zero).

\paragraph{\texttt{renaming} processors}

Rename post-processors can be used to rename parameters. The syntax,
somewhat similar to \texttt{rounding}'s compact one, is the following:

\begin{lstlisting}
{
    "type": "renaming",
    "rename": {
        "old_1": "new_1",
        "old_2": "new_2",
        ...
    }
}
\end{lstlisting}

\noindent
Where \texttt{old\_$k$} are the original name of the parameters we want to
rename and \texttt{new\_k} are the new ones. Note that unlike
\texttt{rounding}'s compact syntax, here \texttt{rename} is an object,
not an array. Also, \texttt{old\_k} are plain strings, not regular
expressions.

\subsection{The \texttt{j2r} command line tool}

All of \textbf{json2run} functionalities are accessed through a command
line utility called \texttt{j2r}. The tool comes with a (large) number
of options, activated by \texttt{-{}-} followed by their long names, or
equivalently \texttt{-} followed by their short names. Some of them have
default values, some other must be provided. (See \texttt{j2r -{}-help}
for a summary of them.)

\subsubsection{Input}

Most of the functionalities of \textbf{json2run} require that you
provide an input file, i.e.~a JSON file describing experiments. This
file is specified through the \texttt{-{}-input} (or \texttt{-i})
option.

\begin{lstlisting}
$ j2r -i experiments.json
\end{lstlisting}

\subsubsection{Available actions}

The main switch in \texttt{j2r} is the \texttt{-{}-action} (or
\texttt{-a}) option. Actions allow to specify what you want
\textbf{json2run} to do for you. The available actions are:

\begin{itemize}
\item
  \texttt{print-cll} prints the generated experiments as a \emph{command
  line list} (also, the default option)
\item
  \texttt{print-csv} print the generated experiments as a CSV file
\item
  \texttt{run-batch} start (or resumes) a batch for the experiments
  described in the input file
\item
  \texttt{run-race} starts (or resumes) a race among parameter
  configurations in order to find the best parameter assignment for the
  specified executable
\item
  \texttt{list-batches} list the batches on the database, and give
  summary information about their completion, machine on which they are
  being run, type of batch, etc.
\item
  \texttt{delete-batch} delete a batch from the database
\item
  \texttt{batch-info} provides detailed information about a batch or
  race, e.g.~repetitions completed, configurations that are still
  racing, experiment file, etc.
\item
  \texttt{rename-batch} rename a batch on the database
\item
  \texttt{show-winning} show winning configurations in a race
\item
  \texttt{set-repetitions} set the number of repetitions of the same
  experiments in a batch or a race
\item
  \texttt{dump-experiments} dump all the experiments data regarding a
  batch as a CSV file
\item
  \texttt{mark-unfinished} set a batch as unfinished, in order to
  restart it
\end{itemize}

\subsubsection{Batch options}

Some of the actions require additional parameters, in particular, each
action pertaining a batch on the database must also provide a mandatory
\texttt{-{}-batch-name} (or \texttt{-n}) to refer it (see it as a key
for batches in the database).

\subsubsection{Running options}

Both \texttt{run-batch} and \texttt{run-race} have a number of
additional parameters that tune the way in which the experiments are
run.

\begin{itemize}
\item
  \texttt{-{}-executable} (or \texttt{-e}) specifies the executable to
  be run (mandatory)
\item
  \texttt{-{}-parallel-threads} (or \texttt{-p}) specifies the maximum
  number of parallel processors to run the experiments onto (defaults to
  the number of cores on the machine where the experiments are run)
\item
  \texttt{-{}-repetitions} (or \texttt{-r}) number of repetitions of the
  (exactly) same experiment to run (e.g.~to have a more reliable result)
\item
  \texttt{-{}-greedy} (or \texttt{-g}) can be \texttt{true} or
  \texttt{false} and states if the batch or race can reuse experiments
  which are already on the database (but are possibly part of other
  batches and races)
\end{itemize}

\subsubsection{Extra options for races}

When a race is run, a number of additional parameters must or can be
passed.

\begin{itemize}
\item
  \texttt{-{}-instance-param} (or \texttt{-ip}) specifies the parameter
  which represents the instance
\item
  \texttt{-{}-performance-param} (or \texttt{-pp}) specifies the
  statistic (output by the executable), that must be used to evaluate
  the quality of a configuration
\item
  \texttt{-{}-seed} (or \texttt{-s}) seed to use to shuffle instances
  (defaults to 0)
\item
  \texttt{-{}-confidence} confidence level for hypothesis testing
  (e.g.~to compare with p-values, defaults to \emph{0.05})
\end{itemize}

\textbf{Note} \textbf{json2run} assumes that the executable outputs a valid JSON
code, with a field for each statistic that we want to record. For
instance, a solver could have a \texttt{cost} and a \texttt{time}
statistic.

\begin{lstlisting}
{
    "cost": 161.12,
    "time": 500
}
\end{lstlisting}

\subsubsection{Database options}

When we're running batches or databases, we're implicitly assuming that
we have a running and accessible MongoDB database. By default,
\textbf{json2run} will look for MongoDB on the \texttt{localhost} and
will try to connect to the database \texttt{j2r} with username
\texttt{j2r} and password \texttt{j2r}. These are just convenient
credentials, but one can specify its own connection parameters through
the following options.

\begin{itemize}
\item
  \texttt{-{}-db-host} (or \texttt{-dh}) specifies the host onto which
  the MongoDB instance is running
\item
  \texttt{-{}-db-database} (or \texttt{-dd}) specifies the database to
  use connecting
\item
  \texttt{-{}-db-user} (or \texttt{-du}) specifies the username to use
  for connecting
\item
  \texttt{-{}-db-pass} (or \texttt{-dx}) specifies the password to use
  for connecting
\item
  \texttt{-{}-db-port} (or \texttt{-dp}) specifies the port to use for
  connecting
\end{itemize}

\subsubsection{Logging info}

By default \texttt{j2r} prints on the standard output most of its
logging information. However this information can be redirected on a
file if needed, and the log level can be set.

\begin{itemize}
\item
  \texttt{-{}-log-file} specifies the file where the log is written
  (default: None)
\item
  \texttt{-{}-log-level} can be \texttt{warning}, \texttt{error}, `info
\end{itemize}

\subsubsection{Source code versioning}

Additionally, \textbf{json2run} can record the code revision used for
running a batch or a race. To enable this option one must pass the name
of the source code manager of choice through the \texttt{-{}-scm} option
(currently supports \texttt{git} and \texttt{mercurial}).

\subsubsection{Instances and configurations}

Instances and parameter configurations are described in the same
experiments file.

\subsection{Running examples}

Here are some of the most common operations that one can perform with
\textbf{json2run}.

\subsubsection{Running a batch of experiments}

Run a batch of experiments based on an experiment file
(\emph{experiments.json}) and an executable (\emph{solver}), with 10
repetitions for each experiment and all the available cores.

\begin{scriptsize}
\begin{lstlisting}
$ j2r -a run-batch -r 10 -n my_batch -i experiments.json -e ./solver
\end{lstlisting}
\end{scriptsize}

\subsubsection{Running a configuration race}

Based on the same file, and reckoning that the instance parameter is
called \emph{instance}, we can run a race to find out the best
configuration. Suppose that the solver outputs some statistics in JSON
(as in the example above) and that we want to compare the configurations
based on the \emph{cost} of the obtained solutions.

\begin{scriptsize}
\begin{lstlisting}
$ j2r -a run-race -r 10 -n my_race -i experiments.json -ip instance -pp cost -e ./solver
\end{lstlisting}
\end{scriptsize}

\subsubsection{Resuming a batch or a race}

To resume a previously stopped batch or race, it is sufficient to run

\begin{lstlisting}
$ j2r -a run-batch -n my_batch
\end{lstlisting}

\noindent
or

\begin{lstlisting}
$ j2r -a run-race -n my_race
\end{lstlisting}

\subsubsection{Printing detailed data about a batch or race}

Use the \texttt{batch-info} action, passing the name of the race or
batch.

\begin{lstlisting}
$ j2r -a batch-info -n my_race
\end{lstlisting}

\noindent
the output is in JSON format (for easy parsing by other tools).

\subsubsection{Print the list of winning (so far) configurations in a race}

Use the \texttt{show-winning} action, passing the name of the race or
batch.

\begin{lstlisting}
$ j2r -a show-winning -n my_race
\end{lstlisting}

\subsubsection{Delete a batch or a race from the database}

Use the \texttt{delete-batch} action, passing the name of the race or
batch.

\begin{lstlisting}
$ j2r -a delete-batch -n my_race
\end{lstlisting}

\subsubsection{List all the batches on the database}

Use the \texttt{list-batches} action.

\begin{lstlisting}
$ j2r -a list-batches
\end{lstlisting}

\subsection{Analyzing the outcome}

The outcome of a batch or race, i.e.~all the data about the experiments,
can be retrieved from R by loading the R script \texttt{analysis.R} and
using the following functions:

\begin{lstlisting}
source("analysis.R")
connect("host") 

x <- getExperiments("my_race", c("instance"))   
\end{lstlisting}

\noindent
The \texttt{x} data frame will contain a row for each experiment in the
batch or race, with information about whether the configuration was one
of the winning ones (in case of a race).

\section{Future}

A new major version of \textbf{json2run} is in the works. Among the upcoming features are:

\begin{itemize}

    \item launching of experiments on multiple machines,
    \item web-service, i.e. RESTful, infrastructure will handle all \textbf{json2run} operations,
    \item improved JSON syntax for all node types (fields and type of values will determine which kind of node are we dealing with), e.g.:
\end{itemize}

\begin{lstlisting}
{
    "type": "and",
    "descendants": [ ... ]
}
\end{lstlisting}

\noindent
will become:

\begin{lstlisting}
{
    "and": [ ... ]
}
\end{lstlisting}

\noindent
and

\begin{small}
\begin{lstlisting}
{
    "type": "discrete",
    "name": "param1"
    "values": { "min": 0.1, "max": 1.0, "step": 0.1 }
}
\end{lstlisting}
\end{small}

\noindent
will become:

\begin{small}
\begin{lstlisting}
{
    "param1": { "min": 0.1, "max": 1.0, "step": 0.1 }
}
\end{lstlisting}
\end{small}

\section{Licensing}

\textbf{json2run} is open-source and distributed under the MIT license.

\section*{Acknowledgements}

\textbf{json2run} has been developed for and with the collaboration of Luca Di Gaspero, Sara Ceschia and Andrea Schaerf of the Scheduling and Time-Tabling Group of University of Udine. Thanks go to Tiago Januario from Universidade Federal de Minas Gerais, Belo Horizonte, Brasil for the many suggestions. Also, thanks go to the \href{http://www.bitbucket.org}{Bitbucket} staff that hosts \textbf{json2run}'s code free of charge.

\bibliographystyle{plain}
\bibliography{paper}

\end{document}